\documentclass[aps,prb,twocolumn,superscriptaddress,overcite,showpacs]{revtex4}
\usepackage{graphicx}
\usepackage{bm}
\def\be{\begin{equation}}
\def\ee{\end{equation}}

\def\mg{\langle g\rangle}

\begin{document}

\title{Intensity distribution of scalar waves propagating in random media}

\author{P. Marko\v{s}} 
\affiliation{Ames Laboratory and Department of Physics and Astronomy, 
Iowa State University, Ames, Iowa 50011}
\affiliation{Institute of Physics, Slovak Academy of Sciences, 
845 11 Bratislava, Slovakia}
\author{C. M. Soukoulis}
\affiliation{Ames Laboratory and Department of Physics and Astronomy, 
Iowa State University, Ames, Iowa 50011}
\affiliation{Research Center of Crete, FORTH, 71110 Heraklion, Crete, Greece}

\begin{abstract}
Transmission of a scalar field through a  random medium,
represented by a system of randomly distributed dielectric cylinders, is
calculated numerically. The  system is mapped to the problem of 
electronic transport  in disordered two-dimensional 
systems. Universality of the statistical distribution
of transmission parameters is analyzed in the metallic  and localized regimes.
In the metallic regime, the  universality of transmission statistics   
in all  transparent channels
is  observed. In the band gaps, we distinguish a 
disorder induced (Anderson) localization from
tunneling through the system, due to a  gap in the density of states. 
We also show  that absorption 
causes a   rapid decrease of the mean conductance, but, contrary to
the case of the localized regime, the conductance is self-averaged with a
 Gaussian distribution.
\end{abstract}


\maketitle
 
\section{Introduction}\label{introduction}

Transport of classical waves in random media is 
a challenging problem, attracting  increasing interest of
theoretical and experimental physicists because  it 
offers the  possibility of  studying Anderson localization.\cite{Anderson}
Since  interactions do not play any role in  classical wave scattering,
these  systems
might be more convenient  for  experimentally verifying  the scaling theory of
localization\cite{AALR} than 
quantum electronic systems, where  the influence of the mutual interaction
 of electrons upon transport has not been yet clarified. \cite{Kravchenko}
 Two main issues   of localization theory, namely
 the presence or  absence of the metallic
state in the  two-dimensional (2D) systems, and the validity of the single parameter 
scaling (SPS)\cite{AALR,MKK,SMO}
might be  more readily resolved experimentally 
 for the classical wave problem than for the electronic one.
Recent experimental results for the transmission of electromagnetic 
waves indeed confirmed that transmission is universal in the diffusive
regime, \cite{SG} and presented strong indications for disordered induced Anderson 
localization. \cite{GG, ChG,ChSG}

In this paper, we analyze numerically the transmission of scalar 
classical waves through 
a two dimensional (2D)  system of randomly distributed dielectric  cylinders.  
Following Ref. \onlinecite{SEGC},
we map the problem into the 2D Anderson model with random binary potential. 
Statistical properties of wave transmission   
are then analyzed using the transfer matrix method. \cite{Ando}
We calculate the conductance $g$  as\cite{Land,ES}
\be\label{land}
g=\sum_{ab} T_{ab}.
\ee
In Eq. (\ref{land}), 
$T_{ab}=|t_{ab}|^2$, where $t_{ab}$ is the transmission 
amplitude from channel $a$ to channel $b$. $a,b=1,2\dots N_{\rm op}$, where
$N_{\rm op}$ is the number of open channels.
We first determine the band structure of the original classical wave problem. Then,
we analyze the statistical properties of the transmission in bands and  gaps.

In bands, where  $g> 1$, we observed diffusive  transport.
Statistical properties of the transmission are in  good agreement
with theoretical predictions of the 
random matrix theory \cite{been,RMP} and the DMPK equation. \cite{DMPK,MS}
The distribution of the conductance is Gaussian with a universal
dimension-dependent variance. \cite{LSF,been,Imry,RMS,Trav}   
Universal properties were predicted  not only for the conductance $g$,
but also for the normalized parameters $s_{ab}=T_{ab}/\langle T_{ab}\rangle$
\cite{kogan,yamilov} and
for the normalized transmission in a given transport channel
\be\label{sa}
s_a=\frac{T_a}{\langle T_a\rangle},\quad\quad T_a=\sum_b T_{ab}.
\ee
The universal probability distribution 
\be\label{dist}
\begin{array}{ll}
&p(s_a)=\displaystyle{\int_{-i\infty}^{+i\infty}\frac{dx}{2\pi}}\exp[xs_a-\Phi(x)]\\
& \\
&\Phi(x)=\mg\ln^2\left(\sqrt{1+x/\mg}+\sqrt{x/\mg}\right)
\end{array} 
\ee
has been  derived analytically.
\cite{NR,Kogan}  
From Eq. (\ref{dist}), the second cumulant  is obtained  as
\be\label{s2c}
{\rm var} s_a=\langle s_a^2\rangle - \langle s_a\rangle^2=\frac{2}{3\langle g\rangle}.
\ee
Universality of the statistical properties of parameters, $s_a$, was  
confirmed experimentally\cite{SG} up to 
rather small values of the conductance ($\mg\approx 2-3$).

In gaps, the mean of the logarithm $\langle\ln g\rangle$ decreases linearly with the system
size. Here, we distinguish  between two different  
regimes, one  with non-zero
density of states, and the other  called tunneling regime, characteristic for the 
frequency region without
eigenstates.  The first describes  Anderson localization, characterized by 
the  Gaussian distribution of $\langle\ln g\rangle$ 
with variance var~$\langle\ln g\rangle\propto -L/\xi$ 
($\xi$ is the localization length) in agreement with localization theory.
The second regime  appears in the
gaps, where the density of states is very small. Then, the  transmission 
is determined  by 
tunneling through the sample. Although $\langle\ln g\rangle\propto - L$,
the distribution of $\langle\ln g\rangle$ is not Gaussian but given by the statistics
of the energy spectra.

The form of the probability distribution $p(\ln g)$ enables 
us also to distinguish between  localization and absorption. 
We find that  absorption also gives a  decrease of $\langle\ln g\rangle\propto -L$; 
however,
in contrast to localization, the conductance remains self-averaged. Our data agree with 
the theoretical prediction, \cite{Brouwer} as well as with the qualitative criterion 
\cite{ChSG}
for localization $2/(3 {\rm  var} s_a)\le 1$.

\medskip

The paper is organized as follows: In Sect. \ref{model}, we introduce the model and system parameters.
In Sect. \ref{map}, we present the mapping of the classical wave problem into the problem of the
transmission of electrons in disordered systems. \cite{SEGC} Results of numerical
simulations are presented in Sect. \ref{four} and \ref{five}. Conclusions are given in Sect. \ref{conc}.

\begin{figure}[t!]
\includegraphics[clip,width=0.4\textwidth]{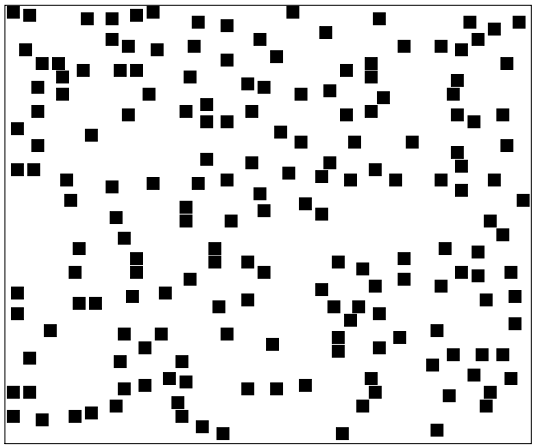}
\caption{Typical sample configuration. The size of the 
system is $128\times 128$, the filling factor $x=0.1$.
182 rectangular rods  with square  cross-section of $3\times 3$
are  randomly distributed throughout the sample. Scalar wave
propagates from left to the right.}
\label{dist-2}
\end{figure}

\section{Model}\label{model}

We study  a two-dimensional system consisting of a random array of 
dielectric cylinders. To make the numerical simulations easier, we
consider rectangular rods
instead of cylinders.
A typical sample  is shown in Fig. \ref{dist-2}. 
No contact or overlap of neighboring rods is  allowed. 
The dielectric permittivity of rods is $\epsilon_2$ and of the  embedding medium 
is $\epsilon_1$.
Two semi-infinite leads
with permittivity $\epsilon_{\rm lead}$ are attached to the sample.
The concentration of the  dielectric rods is given by the
filling factor $x$ of the rod's material.
In this work we consider  square cross-section of dielectric  
$3\times 3$ rods, measured in dimensionless units. 
In the same units, 
the system size $L$ varies  from 32 up to  256.

\section{Mapping to the electronic system}
\label{map}

We use the formal equivalence \cite{SEGC} of the
scalar wave equation,
\be\label{wave}
\nabla^2 u + \epsilon(\vec{r})\omega^2 u=0,
\ee
and  the Schr\"odinger equation for electrons,
\be\label{electron}
\nabla^2\Psi+\left[E-V(\vec{r})\right]\Psi =0,
\ee
(we set $2mc^2/\hbar^2=1$ and $c^2=1$).  
For a given space dependence, $\epsilon(\vec{r})$, and given frequency, $\omega$,
one can find a potential $V(\vec{r})$ and energy $E$ such that the solutions
$u$ and $\Psi$ of Eqs. (\ref{wave}) and (\ref{electron}) are identical.
Formal equivalence requires that the identity
\be\label{equivalence}
\omega^2\epsilon(\vec{r}) = E-V(\vec{r})
\ee
must be fulfilled for all  $\vec{r}$. 

Formula (\ref{equivalence}) does not
mean that the  two models described by Eqs. (\ref{wave}) and (\ref{electron}) are equivalent, 
because potential $V(\vec{r})$ depends both on the
frequency, $\omega$, and the energy, $E$.
 Equation (\ref{equivalence}) only means that for a given
scalar wave model defined by $\epsilon(\vec{r})$, we can find 
for each frequency, $\omega$, an
electronic  model with  potential $V(\vec{r})$ and energy $E$,
that the solutions of both models are the same.

Both the energy and  the potential of the electronic model are
fixed by formula (\ref{equivalence}). Changing  the 
frequency, $\omega$,  we obtain
another electronic model, since the  potential $V(\vec{r})$ changes.
This  means that for a given spatial distribution of $\epsilon(\vec{r})$, 
two different frequencies, 
$\omega$, define two different electronic models.

To be more specific, we consider a model of randomly distributed rectangular rods 
discussed in Sect. \ref{model}.
This model can be mapped into   the electronic model with a random 
binary potential.  For a given sample, the spatial distribution of the potential is
identical with the distribution of the permittivity. 
The energy, $E$, as well as two values of the potential, $V_1$  and $V_2$, is determined 
by 
Eq. (\ref{equivalence}) with the following   relations
\begin{eqnarray}
\omega^2\epsilon_{\rm lead} &=& E\label{zero}\\
\omega^2\epsilon_1 &=& E-V_1\label{one}\\
\omega^2\epsilon_2 &=& E-V_2.\label{two}
\end{eqnarray}

From Eqs. (\ref{one}) and (\ref{two}) we easily obtain
\be\label{omega}
\omega^2\epsilon_1=\frac{\delta}{\mu-1},
\ee
where 
\be\label{mu}
\mu=\frac{\epsilon_2}{\epsilon_1}
\ee
and
\be\label{delta}
\delta=V_2-V_1,
\ee
The energy, $E$, is given as
\be\label{energy}
E=V_1+\frac{\delta}{\mu-1},
\ee
and $V_1$ is determined by
\be\label{v1}
V_1=\omega^2(\epsilon_1-\epsilon_{\rm lead}).
\ee

Note that for $\epsilon_{\rm lead}=\epsilon_0=1$ (vacuum in leads), Eqs.
(\ref{zero}), (\ref{one}, and (\ref{two}) can be solved  only when both $V_1$, $V_2>0$
 assuming that $\epsilon_1,~\epsilon_2>1$. 

For simplicity, we consider in this paper the special case,
\be\label{simple}
\epsilon_1=\epsilon_{\rm lead} =1.
\ee
The method is, of course, applicable to any set ($\epsilon_{\rm lead}$, $\epsilon_1$, $\epsilon_2$),
including $\epsilon_{\rm lead}>\epsilon_1$.

Using Eqs. (\ref{v1}) and (\ref{simple}), we have $V_1\equiv 0$. 
Comparison of Eq. (\ref{omega}) and Eq. (\ref{energy}) gives $\omega^2=E$. Finally,
$V_2=-\delta$. We also fix the ratio of two permittivities $\mu=\epsilon_2/\epsilon_1$,
to the value of $\mu=11$.

\medskip

In numerical simulations, we used the discretized 
version of the Schr\"odinger equation, Eq.  (\ref{electron}),
\be\label{electronp}
\Psi_{x,y+1}+ \Psi_{x,y-1}+ \Psi_{x+1,y}+ \Psi_{x-1,y}=
(4-E+V_{xy})\Psi_{xy}.
\ee
Hard wall boundary conditions  were used. The
transfer matrix method\cite{Ando} was used to calculate numerically 
all the parameters, $s_{a}$, and the conductance. 
The number of open channels, $N_{\rm op}\le L $, which enters  in Eq. (\ref{land})
is given by the number of propagating solutions ($k_n$ real) 
of the dispersion relation
\be\label{disp}
2\cos k_n=4-E-2\cos\frac{\pi}{L+1}n,~~~n=1,2,\dots,L.
\ee
Each frequency, $\omega$, defines a corresponding tight-binding Hamiltonian,
for which the transmission is calculated  using standard  numerical procedures
known for the electronic localization problems.\cite{Ando} Statistical ensembles 
of $N_{\rm stat}=10^4$ samples were considered, which assure that
sufficient accuracy of the transmission parameters of interest were obtained.

\begin{figure}[t!]
\includegraphics[clip,width=0.45\textwidth]{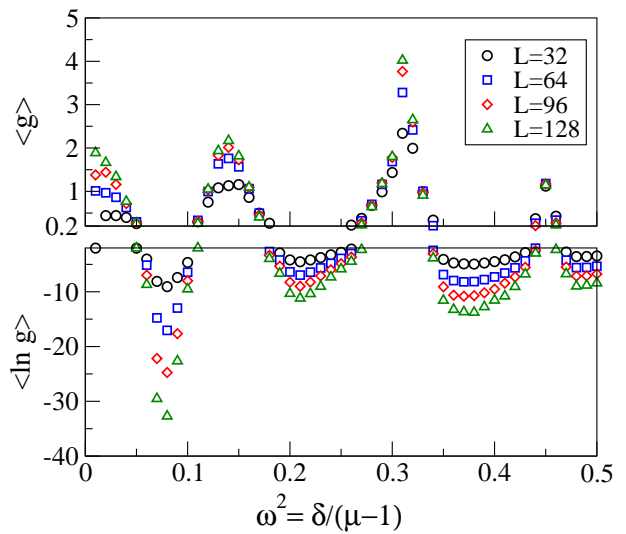}
\caption{Mean conductance $\mg$ (top) and $\langle\ln g\rangle$ (bottom) 
as a function of the frequency for  $\mu=\epsilon_2/\epsilon_1=11$ and
$x=0.2$. Frequency bands and gaps are clearly visible.  
In bands around $\omega^2=0.14$ and 0.31, we find  transport statistics
typical for the  metallic regime. Two different transport regimes
were observed in gaps:
In the first gap, $\omega^2\approx 0.08$,
the density of states is very small and transport 
is due to tunneling through the sample.
In the second gap, $\omega^2\approx 0.21$, we observed disorder 
induced Anderson localization.
}
\label{s2c-02}
\end{figure}

\begin{figure}[t!]
\includegraphics[clip,width=0.45\textwidth]{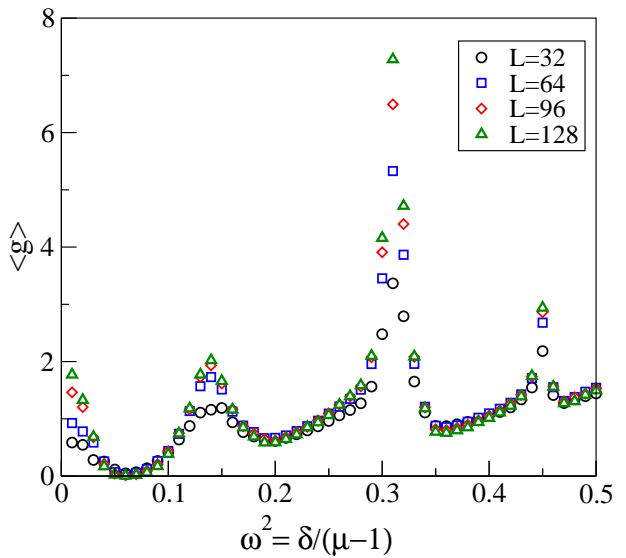}
\caption{The same as in Fig. \ref{s2c-02}, but for $x=0.1$.}
\label{s2c-01}
\end{figure}

\begin{figure}
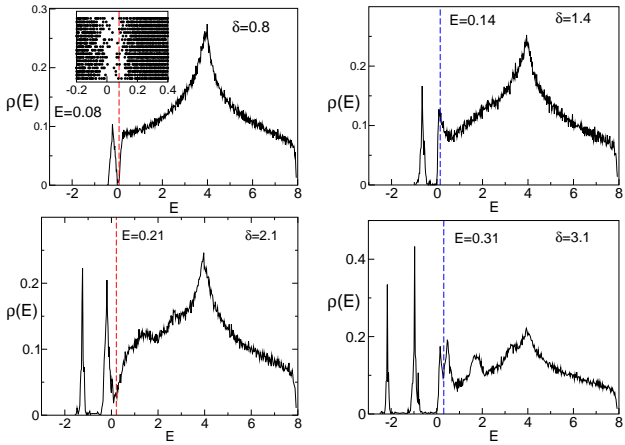

\includegraphics[clip,width=0.22\textwidth]{s2c0_fig4a.eps}~~
\includegraphics[clip,width=0.22\textwidth]{s2c0_fig4b.eps}\\
\includegraphics[clip,width=0.22\textwidth]{s2c0_fig4c.eps}~~
\includegraphics[clip,width=0.22\textwidth]{s2c0_fig4d.eps}
\caption{Density of states of  the electronic system with $x=0.2$ 
calculated for  four 
values of $\delta$.
The size of the system is $48\times 48$ and an average over 10 ensembles was calculated.
Dashed lines indicate the energy, $E(\delta)$. For $\delta=0.8$, the inset  
shows the position of eigenenergies for 18 different samples.}
\label{rho-E}
\end{figure}

\section{Conductance}\label{four}

Figures \ref{s2c-02} and \ref{s2c-01} show the 
$\omega$ dependence of the mean conductance for
the present model. Equivalently, the numerical results of Figs. \ref{s2c-02} and \ref{s2c-01}
might be interpreted as the $\delta$-dependence 
of the conductance of the 2D electronic system. 
In the  last case, however, one must keep in mind that in the Anderson model, the
value of $\delta$ determines the random potential.
Data presented  in Figs. \ref{s2c-02} and \ref{s2c-01}  do not 
directly correspond to the  electronic density of states in the disordered electronic system.
Different values of  $\delta$ correspond to different models.
Note also that the  energy, $E$, Eq. (\ref{energy}) is also 
a function of $\delta$. 

For completeness, we show in Fig. \ref{rho-E} the density of states, $\rho(E)$, 
of  the electronic system for four  values of $\delta$.

\begin{figure}[t!]
\includegraphics[clip,width=0.45\textwidth]{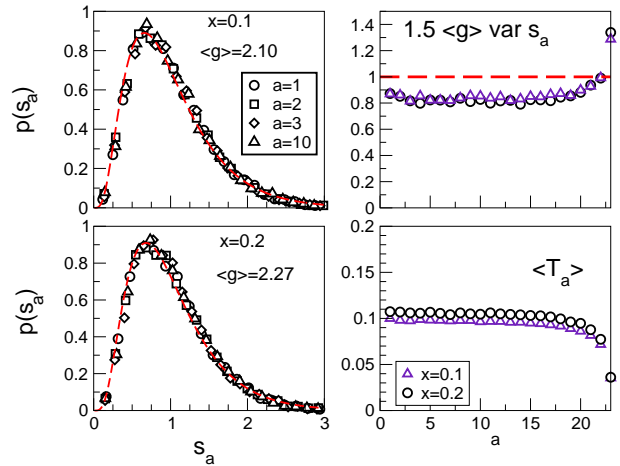}
\caption{Left panels: Probability distribution  $p(s_a)$ for
some open channels   compared with the theoretical prediction of Eq. (\ref{dist})
(dashed line).
$\omega^2=0.14$
(the center of the first band). Concentration of cylinders is 
$x=0.1$ (top) and $x=0.2$ (bottom).
Right panels: The second cumulant  $\frac{3}{2}\langle g\rangle {\rm var} s_a$ as a function of 
index $a$ (top) and  mean values $\langle T_a\rangle$ (bottom).
The size of the system is $L=192$, electron energy is $E=0.14$,
the number of open channels, $N_{\rm op}=23$. Statistical ensemble of
$N_{\rm stat}=10000$ samples was considered.
}
\label{s2c128-d14}
\end{figure}

\begin{figure}[t!] 
\includegraphics[clip,width=0.45\textwidth]{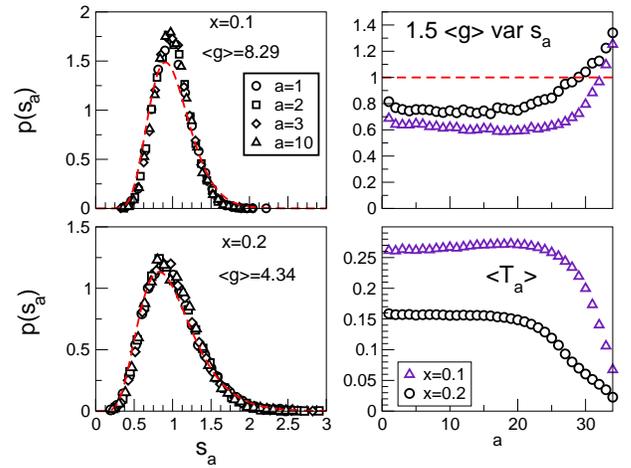}
\caption{The same as in Fig. \ref{s2c128-d14}, but for
$\omega^2=0.31$. $N_{\rm op}=34$. The agreement with theory is not as good as
in Fig. \ref{s2c128-d14}, especially
for $x=0.1$ because of the small system size.
}
\label{s2c128-d31}
\end{figure}

\begin{figure}[t!] 
\includegraphics[clip,width=0.45\textwidth]{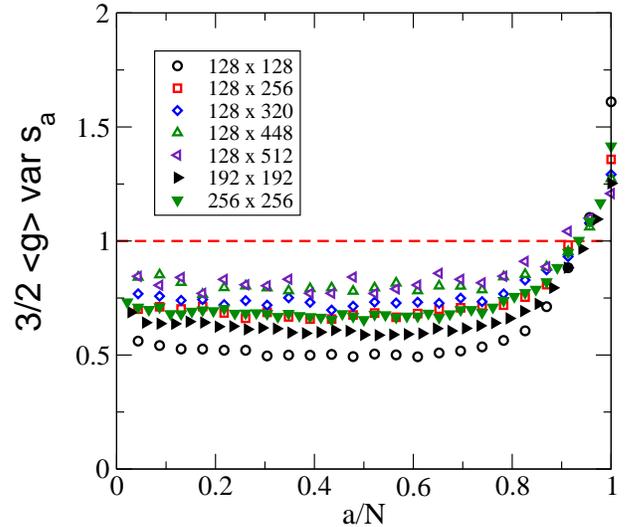}
\caption{The second cumulant $\frac{3}{2}\langle g\rangle \textrm{var} s_a$ as a function
of index $a$ for the center of the second band $\omega^2=0.31$ and $x=0.1$.
2D systems of size $L\times L$ with $L=125$, 192 and 256 were considered.
These  results are  compared  
with quasi-one dimensional systems of size $128 \times L_z$ with $L_z=128$, 256, 
320, 448 and 512
with mean conductance $\langle g\rangle = 7.28$, 4.29,  3.53, 2.57 and 2.25, respectively.
}
\label{s2c128-q1d}
\end{figure}

\subsection{Diffusive regime ($\mg> 1$)}

\begin{figure}[t!]
\includegraphics[clip,width=0.35\textwidth]{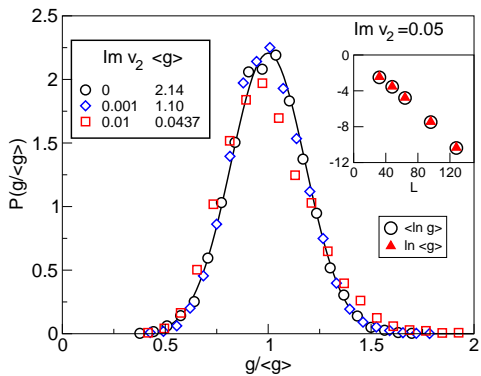}
\caption{The probability distribution of the conductance $g$ for $x=0.2$
and $\omega^2=0.14$ for the case without absorption and with
small absorption in the dielectric cylinders ($L=128$). 
 Absorption is
due to the small imaginary part of the permittivity $\epsilon_2$ of cylinders.
The mean conductance decreases,  due to  absorption (see legend) and 
the conductance  distribution is  Gaussian. 
The inset shows the $L$ dependence of $\ln \langle g\rangle$ and $\langle\ln g \rangle$
for the stronger absorption (Im $V_2=0.05$). The numerical results
 confirm that the conductance is
self-averaged. 
}
\label{s2c128-02-d14-g}
\end{figure}

Three frequency bands are visible in Figs. \ref{s2c-02} and \ref{s2c-01},
where  we  expect the 
metallic behavior. Mean conductance $\langle g\rangle >1$ slightly 
increases with the system size
and the  value of var $g$ is very close to the universal conductance fluctuation.
\cite{LSF,RMS,Trav} 
  The distribution 
of the conductance is Gaussian (see, for example, Fig. \ref{s2c128-02-d14-g}). 
Although these properties are
finite size effects (no metallic state exists in 2D in the limit $L\to\infty$), 
the numerical data 
enable us to check the theoretical prediction about the transmission statistics.

In Fig. \ref{s2c128-d14} we show the statistics of the parameters, 
$s_a$,
for the frequency in the center of the first ($\delta\approx 1.4$) band. 
The number of open channels is 23 
for the size  of the system $192\times 192$. Results confirm that the distribution,
$P(s_a)$, is universal and does not depend on $a$ . The second cumulant, 
${\rm var} s_a$, is close to its theoretical value, $2/3\mg$.
 
\begin{figure}
\includegraphics[clip,width=0.42\textwidth]{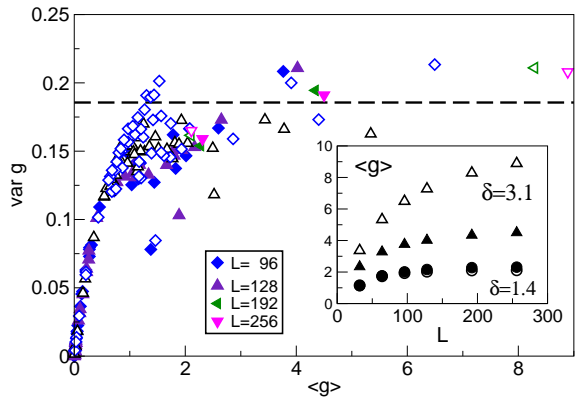}
\caption{The variance of the conductance, var $g=\langle g^2\rangle -\langle g\rangle^2$,
as a function of the mean conductance for $x=0.1$ and $x=0.2$, and various system sizes.
Open symbols: $x=0.1$, full symbols: $x=0.2$.
For small values of $\langle g\rangle$, the numerical results  scale 
to a universal curve. 
For large $\langle g\rangle$ they
converge to the universal  value of var~$g=0.1855$. \cite{LSF,RMS}
The inset shows the mean conductance $\langle g\rangle$ 
for $\delta=1.4$ (circles) and $\delta=3.1$
(triangles). For $\delta=3.1$,  $\langle g\rangle$ still 
increases with $L$, which indicates
that $L$ is not large enough  for transport to be  diffusive. 
}
\label{univ}
\end{figure}

The same analysis was completed  for the second band (Fig. \ref{s2c128-d31}). 
Here, the  agreement with theory is not as good as in the previous case, 
presented in Fig. \ref{s2c128-d14}  especially for a smaller concentration
of rods.  Although the distribution, $P(s_a)$, does not depend on $a$, 
it differs considerably
from theoretical predictions.  We interpret this discrepancy as a finite size effect.
Indeed, the mean conductance $\langle g\rangle$ increases with the system size
(Fig. \ref{s2c-01}, and inset of Fig. \ref{univ}), which indicates that we have not
reached the diffusive regime yet.

Note also that  we are studying
2D samples,  while the theory is formulated for quasi one dimensional systems. 
Therefore, we expect that the agreement with  theory should be better if the
length of the system increases.  This is confirmed by numerical results
presented in Fig. \ref{s2c128-q1d},
which shows the second cumulant
$\frac{3}{2}\langle g\rangle \textrm{var}~s_a$ for various 2D and quasi one dimensional system.

Figure \ref{univ} shows the variance, var~$g$, of the conductance {\sl vs}  
the mean conductance $\langle g\rangle$.
We expect that 
${\rm var}~g$ should  saturate to the
universal value  of var~$g\to 0.1855$ [\onlinecite{LSF,RMS}] for $g\gg 1$.

\begin{figure}[t!]
\includegraphics[clip,width=0.35\textwidth]{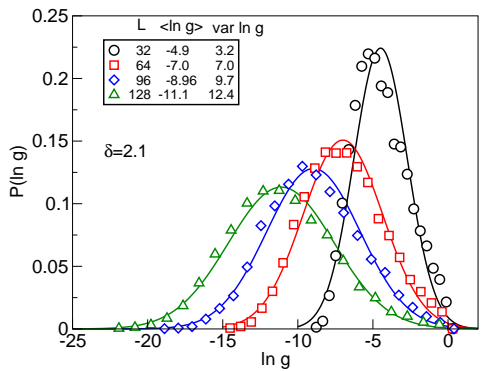}
\caption{The probability distribution $P(\ln g)$ for
$\omega^2=0.21$. The distribution is close to the log normal
with var $\ln g\approx -\langle\ln g\rangle$, 
typical behavior of Anderson localization. 
The number of random configurations 
$N_{\rm stat}$ is $10^5$ for $L\le 96$ and $10^4$ for $L=128$. The legend presents 
the data for $\langle\ln g\rangle$ and  var $\ln g$. 
}
\label{s2c-lng-d21}
\end{figure}

\begin{figure}[t!]
\includegraphics[clip,width=0.35\textwidth]{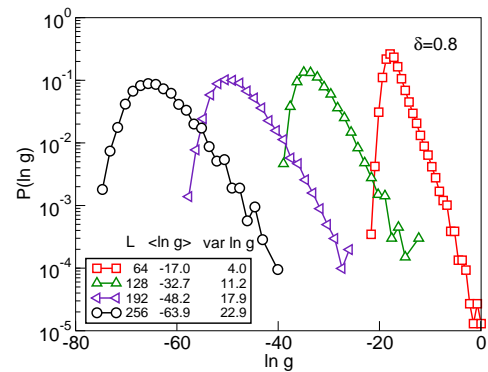}
\caption{The probability distribution  $p(\ln g)$ for 
$\omega^2=0.08$ and $x=0.2$. The distribution shows that transport is due to 
tunneling through a system with an energy gap. Note that the width
of the distribution, var $\ln g\ll\langle\ln g\rangle$,
 is much smaller than that of the disorder induced insulator.
Data for the mean value and variance of ln $g$  are given  in the legend. 
$N_{\rm stat}=7000$ for $L\le 256$ and is $>10^4$ for smaller $L$.
}
\label{s2c-lng-d08}
\end{figure}

\subsection{Localization  ($\mg\ll 1$)}

In the regions between the pass bands, the conductance 
decreases exponentially with the system size.
It is much more pronounced for larger  concentrations of cylinders 
($x=0.2$). Here, two different regimes were
observed. 

In the upper gap, ($\delta\approx 0.21$) the  observed statistical properties
of the conductance  are in agreement with the
theoretical expectations for the localized regime: 
The distribution of $\ln g$ is Gaussian (Fig. \ref{s2c-lng-d21}) with  
var $\ln g\approx -\langle\ln g\rangle\propto 2L/\xi$, which is characteristic 
for the disorder induced localization. The parameter $\xi$ is  the localization length.
\cite{pozn}

In the lower gap ($\delta\approx 0.08$), 
the conductance decreases rapidly as the size of the
system increases. The probability distribution of $\ln g$ is not Gaussian,
as can be seen in Fig. \ref{s2c-lng-d08}.
 Instead,
 it  decreases exponentially  for larger values 
of conductance as
\be
p(\ln g)\propto \exp \textrm{const}[\langle\ln g\rangle - \ln g].
\ee
In contrast to Anderson localization, no
samples with conductance close to 1 were  found. 
This indicates that transport 
is possible only by tunneling through isolated eigenstates.
Also, the variance, var~ln~$g$, is much smaller than that of  the Anderson insulator,
\be
\textrm{var} \ln g\ll -\langle\ln g\rangle.
\ee
The last property seems to be in agreement with previous work of 
Deych \textsl{et al.} \cite{deych} who argued that single parameter 
scaling does not work in the
energy intervals, where the density of states is so small that   
another characteristic length
$l_s\sim\sin^{-1}\rho(E)$ exceeds the localization length. 
Indeed, we  found that the density of states is close to zero 
in the neighborhood of $E=0.08$
(Fig. \ref{rho-E}),
so that the average distance between isolated eigenstates is 
larger than the localization length ($\xi\approx 8$ was estimated form the 
$L$ - dependence $\langle\ln g\rangle$). 
For instance, in the interval $0.05<E<0.11$,  we found an average of only one eigenstate, 
when  $L=48$ so that $l_s>48$.  Although $l_s$ is expected to decrease
when $L$ increases, we were not able to reach Anderson localization, 
even for the largest  system studied,  $L=256$.

\section{Absorption}\label{five}

Absorption reduces the transmission of the EM waves in a similar way as
localization. Mean conductance decreases exponentially with the 
system length.\cite{Brouwer} 
To  distinguish between  Anderson localization
and absorption effects, we need to understand the statistical properties 
of the transmission. In Ref.   \onlinecite{ChSG} the simple criterion for localization 
was derived, based on   the value of the parameter
\begin{equation}\label{gprime}
g'=2/(3 {\rm var} s_a). 
\end{equation}
It was argued that localization appears if $g'\le 1$.

To study the effects of  absorption, we add a small imaginary part to the
permittivity of cylinders (more precisely, $V_2$ in Eq. (\ref{two})
 becomes complex in our simulations).    
First, we analyze how absorption changes the transmission 
properties of the metallic system.
We use $\delta=1.4$ and $x=0.2$ (Fig. \ref{s2c-02}).
As expected, the   mean conductance decreases 
when the system size increases, $\langle\ln g \rangle\sim -L$ 
similarly as for  localized waves.
In contrast to the localized regime, the
conductance is  self-averaged in this case, 
$\langle\ln g\rangle=\ln\langle g\rangle$. 
The conductance distribution is still  Gaussian. 
As shown in Fig.  \ref{s2c128-02-d14-g}, the width of the distribution of
the \textsl{normalized} conductance depends only weakly on the absorption strength.
This is in agreement with analytical results of Brouwer. \cite{Brouwer}

\begin{figure}[t!]
\includegraphics[clip,width=0.45\textwidth]{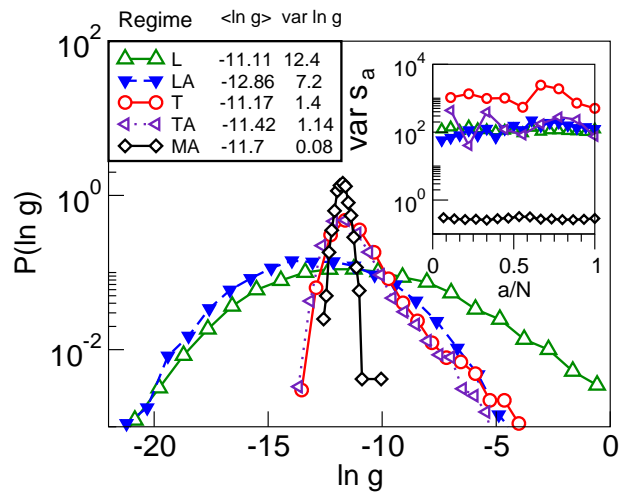}
\caption{The probability distribution of the logarithm of the conductance in various regimes.
L: regime of Anderson localization with Gaussian distribution and 
${\rm var} \ln g\approx -\langle\ln g\rangle$.
LA: localized regime with absorption (Im $V_2=0.01$). Compared with L, we see that
the part of the
distribution with relatively large conductance is missing.
T: tunneling regime, in which 
${\rm var} \ln g \ll -\langle\ln g\rangle$.
TA: tunneling with absorption (Im $V_2=0.05$). Absorption does not influence $p(\ln g)$.
MA: metallic regime with absorption (Im $V_2=0.065$). 
Here, conductance is self-averaged.
Parameters of systems  were chosen such that 
$\langle\ln g\rangle$ is approximately
the same for all systems (see legend).
Inset shows var $s_a$ in all the  above regimes.  
As predicted in Ref. \onlinecite{ChSG},
var $s_a<1$ in the metallic regime with absorption, 
but is $\gg 1$ in the localized regime,
both with and without absorption. The same holds for the tunneling regime
(data for L, LA and TA regimes are almost indistinguishable). Thus,
the criterion $g'<1$ cannot distinguish between localization and tunneling.}
\label{porovnanie1}
\end{figure}

Figure \ref{porovnanie1} compares the statistics of the logarithm of the conductance of
five different transport regimes --  localization and  tunneling  with and without absorption, 
and  diffusive (metallic) regime with absorption.  
The numerical results  confirm that absorption does not change the 
statistical properties  of the conductance,  given mostly by the  interference  of
electrons or classical waves due to disorder. 
In the regime of Anderson localization,  the probability to find relatively large
values of $g$ is reduced due to absorption, while another part of the distribution,
where  $\ln g\ll \langle\ln g\rangle$, is almost unaffected by the presence of absorption.
As a result, $p(\ln g)$  is  not Gaussian anymore. 
In the tunneling regime, absorption only reduces the magnitude of var $s_a$
as can be seen in the inset of Fig. \ref{porovnanie1}.

\section{Conclusions}\label{conc}

We presented a numerical analysis of the transmission of the scalar classical wave through
a  disordered two dimensional system. By mapping this system to a 2D electronic disordered
problem, we found frequency intervals  with high transmission. For these frequencies, we
obtained the statistical distribution of the transmission parameters 
predicted recently by the theory.
We confirmed the universality of the conductance fluctuations and of the distribution 
of the parameters $s_a$ in the metallic regime. 
The universality survives for rather small values of conductance,
$\langle g\rangle = 1-3$.  Our numerical results
 confirm that the theory, developed for the
quasi one dimensional systems, can be successfully applied to 2D systems, too. 
Our results are also  consistent with previous experiments. \cite{SG}

As there is no metallic regime in two dimensional systems, 
the above described metallic
behavior is just an effect of the finite size of our sample. By increasing the system size,
conductance would decrease and finally, in the limit of $L\gg$ localization 
length, the  wave becomes
localized. The localized regime was also  observed  in a gap, where the density of states
is smaller than in the bands, but still non-zero. In this frequency interval, 
broad distribution of the logarithm of the conductance,
typical for the Anderson localization in electronic systems, is observed.

Anderson localization should be distinguish from the  tunneling regime,  which we found
in the frequency gap, where the density of states is close to zero.
Here, the statistics of conductance is determined by the
statistical properties of the isolated frequencies inside the gap. The distribution
$p(\ln g)$ differs considerably from Gaussian. These results are in agreement with
the theoretical analysis of Deych \textsl{et  al.} \cite{deych}

Finally, we analyzed the effects of absorption.  We found, in agreement with
theoretical and experimental  works,  that absorption does not change the
statistical properties of the transmission.  
Since the  statistical properties of the parameters, $s_a$, are 
insensitive to the presence of absorption both in the localized and in the 
metallic regimes,
typical values of var $s_a$  enable us to decide whether the  
exponential decrease of the conductance is
due to localization or  to  absorption. However,
statistics of $s_a$ can not distinguish between Anderson localization and tunneling. 
Fortunately, these two regimes are 
distinguishable from the form of  the probability distribution of the total transmission $g$.

\noindent{\bf Acknowledgments.}
This work was supported by Ames Laboratory (Contract. No. W-7405-Eng-82),
and   VEGA (Project No.  2/3108/2003).

\end{document}